\documentclass{article}
\usepackage{apacite}
\usepackage{graphicx} 
\usepackage{natbib} 
\usepackage{amsmath} 
\usepackage{color}
\usepackage{amssymb}
\usepackage{gensymb}
\usepackage{subfloat} 
\usepackage{caption}
\usepackage{subcaption}
\usepackage{float}
\usepackage{siunitx}
\usepackage{multirow}
\usepackage{booktabs}
\usepackage{hyperref}
\usepackage{diagbox}
\usepackage{xfrac}

\title{\textbf{Bayesian spatial extreme value analysis of maximum temperatures in County Dublin, Ireland}} 

\author{John O'Sullivan, Conor Sweeney, Andrew C. Parnell} 

\date{} 

\begin{document}

\maketitle 

\begin{abstract}
In this study, we begin a comprehensive characterisation of temperature extremes in Ireland for the period 1981-2010. We produce return levels of anomalies of daily maximum temperature extremes for an area over Ireland, for the 30-year period 1981-2010. We employ extreme value theory (EVT) to model the data using the generalised Pareto distribution (GPD) as part of a three-level Bayesian hierarchical model. We use predictive processes in order to solve the computationally difficult problem of modelling data over a very dense spatial field. To our knowledge, this is the first study to combine predictive processes and EVT in this manner. The model is fit using Markov chain Monte Carlo (MCMC) algorithms. Posterior parameter estimates and return level surfaces are produced, in addition to specific site analysis at synoptic stations, including Casement Aerodrome and Dublin Airport. Observational data from the period 2011-2018 is included in this site analysis to determine if there is evidence of a change in the observed extremes. An increase in the frequency of extreme anomalies, but not the severity, is observed for this period. We found that the frequency of observed extreme anomalies from 2011-2018 at the Casement Aerodrome and Phoenix Park synoptic stations exceed the upper bounds of the credible intervals from the model by 20\% and 7\% respectively.
\vspace{15 pt}
 
\end{abstract}


\section{Introduction}

In this study, we produce return-levels of anomalies of daily maximum temperature extremes for an area over Ireland, for the 30-year period of 1981-2010. We apply Bayesian hierarchical modelling \citep{gelman2013bayesian} combined with a reduced-rank method called predictive processes to solve the computationally difficult problem of modelling data over a very dense spatial field \citep{banerjee2008gaussian}. The role of extreme value theory (EVT) \citep{coles2001introduction} is increasingly important in furthering our understanding of climate change and climate extremes. With an increase in maximum temperature extremes through the current century projected by the Intergovernmental Panel on Climate Change (IPCC), it is important to have a better understanding of recent observations of climate extremes \citep{IPCCAR5WG1}. Ireland is ``completely off course in terms of achieving its 2020 and 2030 emission reduction targets", according to the Climate Change Advisory Council's Annual Review 2018. Without urgent action leading to substantial reductions in greenhouse gas emissions, they warn that Ireland is unlikely to deliver on its national, EU and international obligations \citep{CCAC}. A better understanding of our current climate  extremes is essential to enable both the general public and policymakers to plan better in order to mitigate or avoid the many impacts of climate change; whether social, infrastructural, environmental, or economic \citep{IPCCAR5WG3}.
\\

In order to model extremes of daily maximum temperature, the chosen statistical models must use EVT because extremes of temperature are rare (by definition), and occur in the tails of the distribution. Since extremes of temperature vary by location on any given day, the statistical models used should account for this spatial dependence. And since these distributions are governed by parameters that depend not only on the data, but on other (e.g., physical or mathematical) principles and constraints, it is natural to use a Bayesian framework. With this in mind, we apply a Bayesian hierarchical spatial extreme value model to a dataset of daily maximum temperatures in Ireland.
\\

Estimating hierarchial Bayesian models using Markov chain Monte Carlo (MCMC) methods involves matrix factorisations of the order of $n^{3}$, where $n$ is the number of locations \citep{guhaniyogi2011adaptive}. In our case, with a dataset of daily maximum temperature from 1981 to 2010 at more than $\sim$72,000 locations, this becomes computationally infeasible. In order to overcome this problem, we focus our attention on the capital city of Ireland and its surroundings, and employ reduced-rank spatial models. These are very popular models for analysing large spatial datasets \citep{banerjee2008gaussian}. In particular, we work with a flexible class of low-rank models called predictive process models. 
\\

We aim to begin a comprehensive characterisation of temperature extremes in Ireland for the period 1981-2010. Our contribution here expands upon existing research on historical temperature extremes in Ireland, which to our knowledge, consists only of site-specific analysis with no spatial component to the models used (\textit{e.g.}, \citet{walsh2012summary} and \citet{osman2015modelling} - described in the next section). All code needed to reproduce our analysis is available in a public repository on GitHub at \href{https://github.com/jackos13/extremes}{https://github.com/jackos13/extremes}. We note that this repository does not include the data; they are available upon request from Met \'Eireann.
\\

The structure of the rest of the article is as follows. Section 2 reviews the body of literature of previous work in the area. Section 3 describes the dataset and covariates used in this study. Section 4 comprises four subsections describing the methodology: the first subsection provides an overview of EVT; the second provides an overview of standard spatial statistics theory and predictive processes; the third discusses the issue of threshold selection for our data; and in the fourth subsection, we detail our selected model. Section 5 illustrates the main results from the model-fitting stage. Section 6 discusses these results, putting them in context with other studies, and concludes the article.


\section{Previous relevant work in climate extremes and spatial modelling}

\subsection{Extreme climate work in Ireland}

In a study of Ireland's climate from 1981 to 2010, \citet{walsh2012summary} produces a table of monthly values of mean daily maximum temperatures for two synoptic stations, one at Casement Aerodrome and one at Valentia Observatory. At Casement Aerodrome (in Dublin), the annual mean daily maximum temperature is found to be 13.4\degree{C}. The monthly mean daily maximum temperature at this station ranges from 8\degree{C} in January to 19.8\degree{C} in July. A further breakdown by individual days is not provided. \citet{osman2015modelling} investigate temperature extremes at six different locations in Ireland. They fit a generalised Pareto distribution (GPD) to the observed extremes at each location (for the period 1961-2000), allowing the scale parameter to vary as a function of large-scale climate variables derived from reanalysis data. From these models, they then produce return-level plots for all six locations, leading them to conclude that significant changes in extreme temperature events are projected to occur in Ireland over the course of the 21st century. These include hotter summers and milder winters.

\subsection{Spatial modelling for large data sets}

In the context of using Bayesian hierarchical spatial models to analyse the large spatial datasets that occur frequently in environmental sciences, \citet{guhaniyogi2011adaptive} discuss predictive process models. They point out the flexibility of these models, explaining how they can overcome the `big $n$' problem of large datasets, where matrix factorisations of the order of $n^3$ (where $n$ is the number of locations) can make the direct estimation of hierarchical spatial models infeasible. \citet{finley2012bayesian}, using a Bayesian hierarchical framework, apply a predictive process model to mean temperature data from 2000-2005 from weather stations situated across the northeastern U.S.A. They found that the low-dimensional predictive process model was highly effective at borrowing information over space to make accurate and precise predictions for new locations. In addition to this, the choice of a Bayesian framework meant that the authors could determine the full posterior predictive distribution at any location.

\subsection{Spatial extreme value analysis}

In order to produce maps of precipitation return levels and uncertainty measures over a region in Colorado in the U.S.A., \citet{cooley2007bayesian} construct two separate Bayesian hierarchical models for extreme precipitation events: one for the intensity (using the GPD), and another for the frequency (using the Binomial distribution) of such events. The assumption underlying both models is that regional extreme precipitation is driven by a latent spatial process, characterised by geographical and climatological covariates. Their approach involves pooling all of the information from different stations in order to produce a 25-year daily precipitation return level map, which directly takes into account the parameter and interpolation uncertainty in the method itself. \citet{shaby2012bayesian} apply a Bayesian hierarchical spatial extreme value model to temperature extremes across Europe in order to assess the changes in the risk of widespread extremely high temperatures across agricultural land. For their data, they use annual maximum temperatures on a selected subgrid of 985 locations. They find that the risk of large percentages of cropland exceeding a high temperature threshold has probably increased in the last century, but only slightly so. \citet{lehmann2016spatial} use a Bayesian hierarchical framework and a block-maxima approach to model extremes of precipitation of different durations from over 1,300 weather stations in two different regions in Australia. The parameters of the distribution are modelled as spatial Gaussian processes. They found that pooling the data across space, and thus borrowing strength from neighbouring stations, leads to more precise parameter estimates and therefore superior posterior inference. This borrowing of strength is particularly important when dealing with extremes, which are, by definition, rare. In a similar study, \citet{dyrrdal2015bayesian} also used spatial Gaussian processes in a Bayesian hierarchical model in order to produce spatial maps of extreme hourly precipitation over Norway. They specify a generalised extreme value distribution at each point in space, and allow the parameters to depend on location-specific geographic and meteorological variables, a structure similar to generalised linear modelling. Variable uncertainty is accounted for using Bayesian Model Averaging (BMA). They find that their approach performs well at estimating extreme hourly precipitation return levels, both in terms of magnitude and spatial distribution. \citet{ghosh2011hierarchical}, using a Bayesian hierarchical approach, compare two separate data models (a GEV distribution and a variation of the GPD) to model precipitation extremes over continental U.S.A. from 1900 to 1998. They find that the peaks-over-threshold approach (i.e., using the GPD) shows better fit and improved forecast ability to the block-maxima approach (i.e., the GEV).


\section{Data}

\begin{figure}[H]
    \centering
    \begin{subfigure}{.5\textwidth}
  \centering
  \includegraphics[width=1\linewidth]{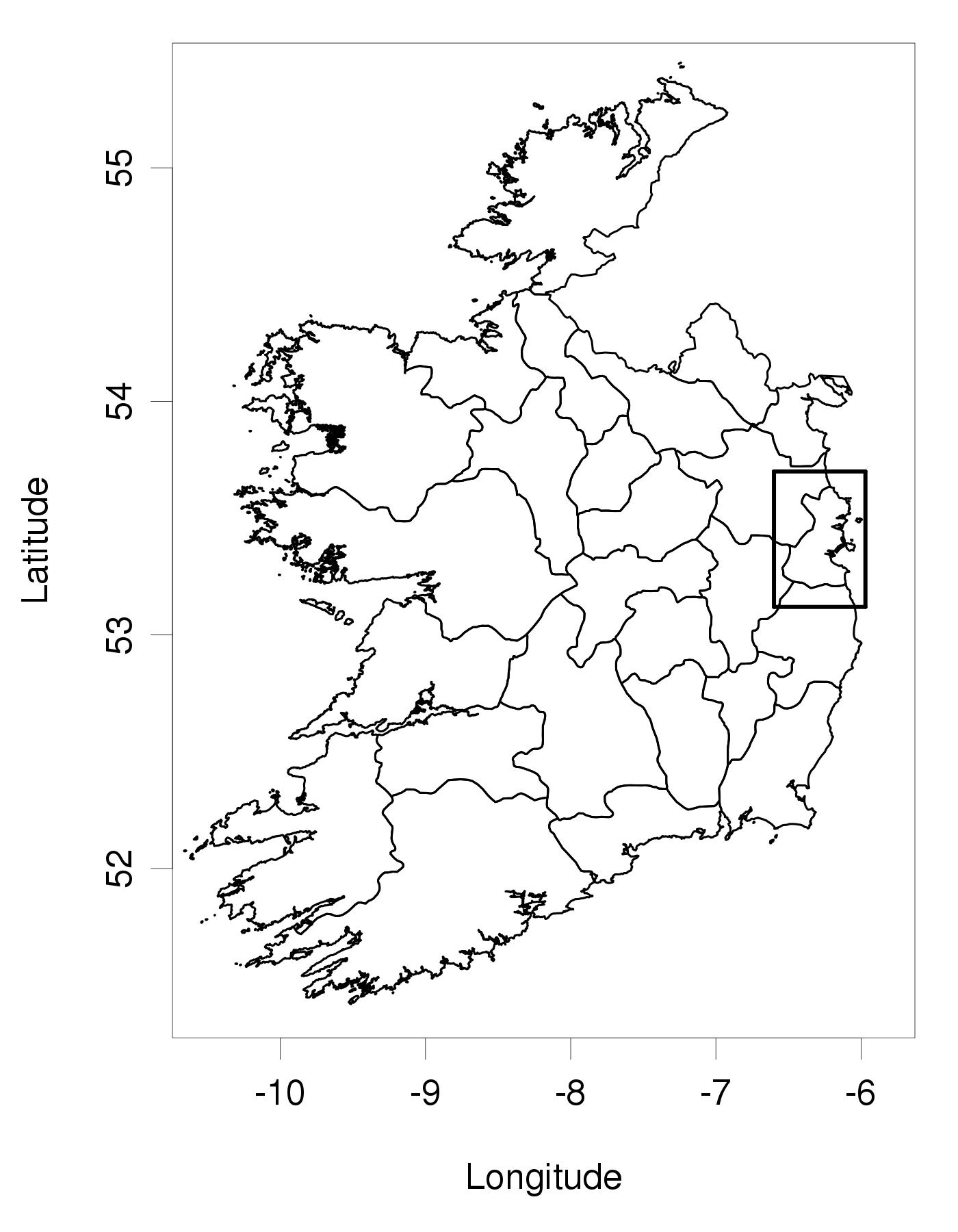}
\end{subfigure}%
\begin{subfigure}{.5\textwidth}
  \centering
  \includegraphics[width=1\linewidth]{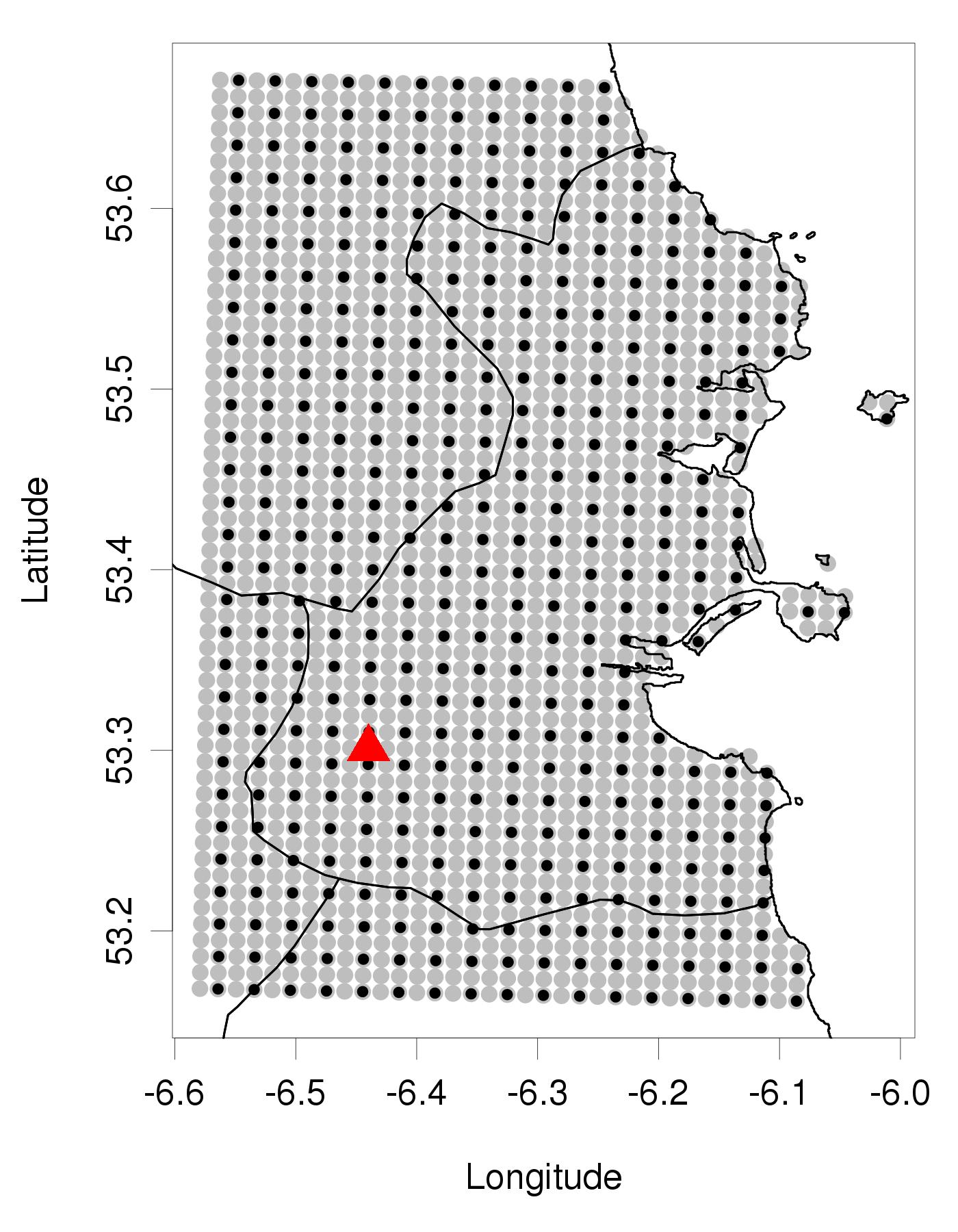}
\end{subfigure}
\caption{The data domain showing the outline of the state of Ireland (with county outlines included) is on the left of the image. The area within the black box is our study region, and is blown-up and shown on the right. Here, the full grid (grey circles) and the selected subgrid (black circles) are illustrated. A red triangle marks the location of Casement Aerodrome, for which we have synoptic station data that is used later in the paper for model evaluation purposes.}
\label{fig:dublin}
\end{figure}

For this study, we make use of a gridded dataset of daily maximum temperatures over Ireland. The data is on a $1 \times 1$ km$^{2}$ grid ($\sim$72,000 gridpoints), consisting of daily values of maximum temperature for the 30-year period 1981-2010. This was produced by Met \'Eireann's Climatology and Observation division, using observational data from 138 stations and independent variables available at each grid point, such as elevation, latitude, and longitude. This observational data was interpolated onto the grid, using an inverse distance weighted algorithm \citep{walsh2017} (available following a suitable request to the Climate Enquiries Office at Met \'Eireann: https://www.met.ie/). Figure \ref{fig:dublin} shows the domain of the dataset. The image on the left illustrates the extent of the full dataset, which covers the state of Ireland. Geographic covariates of latitude, longitude and altitude are available on the same $1 \times 1$ km$^{2}$ grid as the temperature data. The plot on the right of Figure \ref{fig:dublin} shows the region we chose to focus on in our research.

\subsection{Study domain} 

Due to the large number of gridpoints in the dataset, we decided to focus on a smaller section to ensure that running the model was computationally feasible. We focused on an area covering County Dublin, which includes the capital city and its surroundings, located in the east of the island. This study domain is shown in Figure \ref{fig:dublin}(b). It contains approximately 1,700 gridpoints. The full grid is displayed with grey circles and the selected subgrid is shown with black circles. The subgrid is constructed by selecting every second gridpoint in both the horizontal and vertical directions. Also marked here with a red triangle is the location of Casement Aerodrome, where a synoptic meteorological station is situated. This is used in the specific-site analysis later in this section and in section 4.

\subsection{Data verification}

\begin{figure}[H]
    \centering
  \includegraphics[width=0.6\linewidth,keepaspectratio]{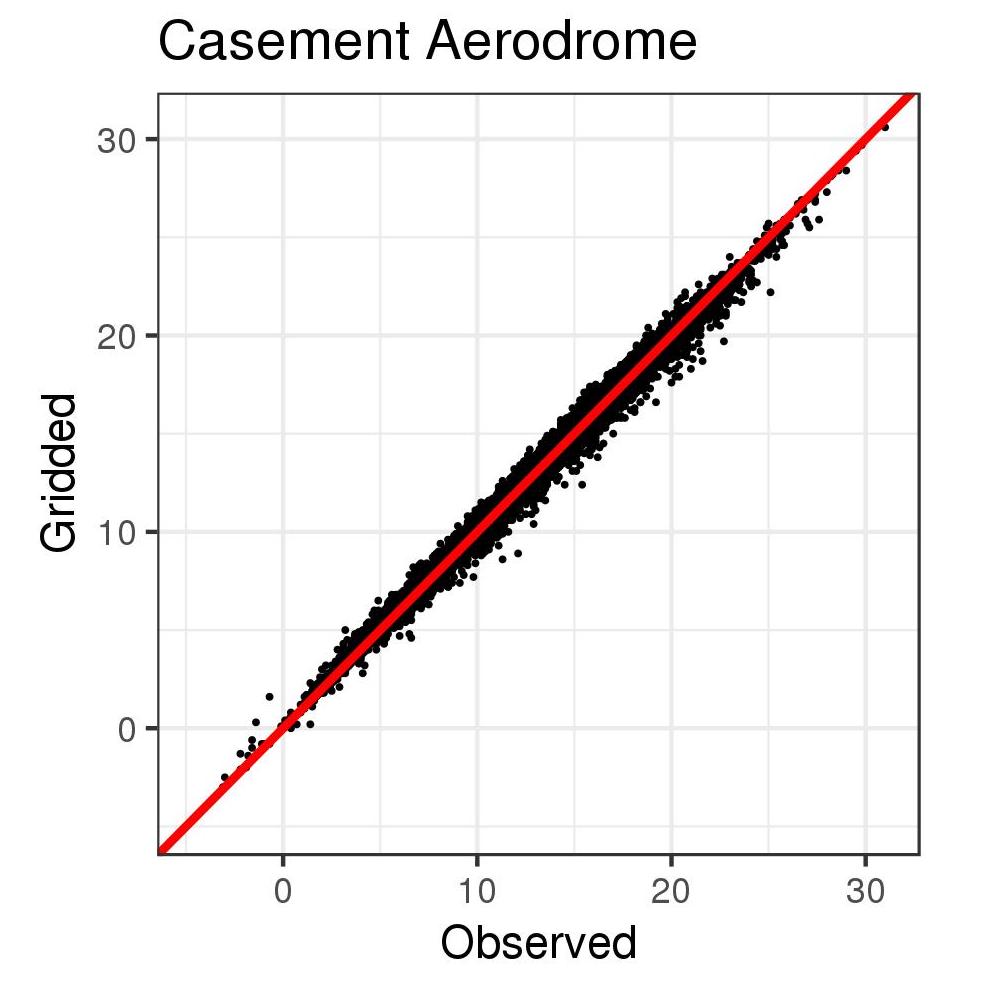}
\caption{Shown is a scatterplot of observed \textit{vs.} gridded temperatures at Casement Aerodrome. A very strong positive correlation between the observed and the gridded data is clearly visible (correlation = 0.997).}
  \label{fig:verify1}
\end{figure}

\begin{figure}[H]
    \centering
  \includegraphics[width=0.8\linewidth,keepaspectratio]{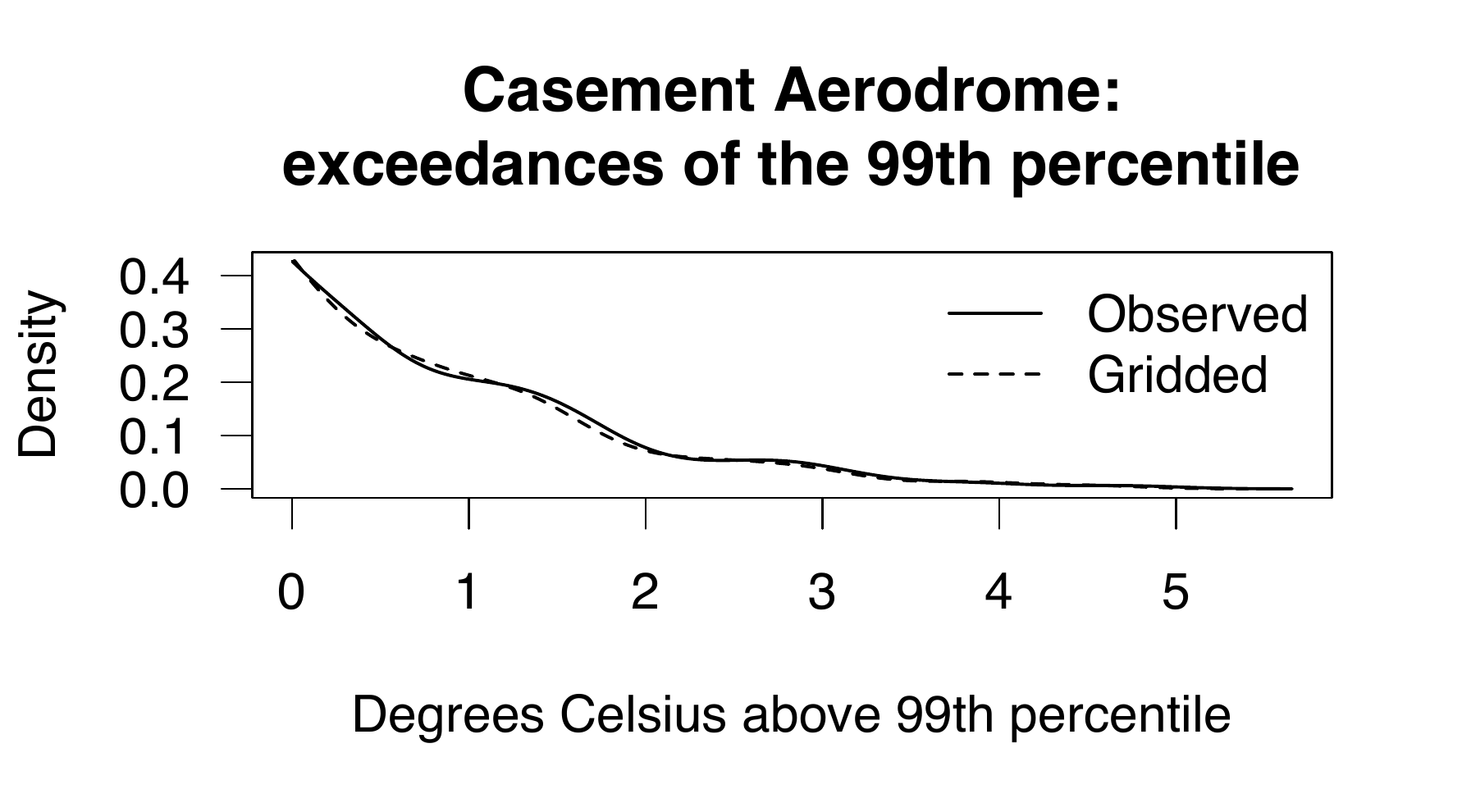}
\caption{Shown are the two overlaid densities (of the daily maximum temperature  gridded data and observations) of the exceedances above the 99th percentile at Casement Aerodrome (area of overlap = 0.95).}
  \label{fig:verify2}
\end{figure}

We performed various checks of the gridded data in order to compare it to the underlying synoptic station observations which were used in generating it. Two of these site-specific checks are illustrated in Figures \ref{fig:verify1} and \ref{fig:verify2}. The observations here are taken from the synoptic station at Casement Aerodrome (location shown in Figure \ref{fig:dublin}(b)), while the gridded data is taken from the nearest gridpoint (approximately 540 m away).
\\

Firstly, we created a scatterplot of observed \textit{vs.} gridded data. This shows (Figure \ref{fig:verify1}) a very strong positive correlation between the observed and the gridded data, as would be expected. At all 9 sites checked across the full domain (i.e., nationwide), there was a correlation of $>0.985$ between the observations at the synoptic station there and the gridded observations at the nearest gridpoint. For the data illustrated here, the correlation is 0.997.

Secondly, in order to focus on the tail of the data (which is where our interest lies), we then considered only the data above the 99th percentile in both cases: observations and gridded data. The second panel (Figure \ref{fig:verify2}) shows these two smoothed densities overlaid (where the respective thresholds were subtracted from the two datasets in order to have them both begin at 0 to aid comparison). A very close relationship can be seen between the two densities: the observations and the gridded data have a very similar shape (area of overlap calculated from the two kernel density estimations for the empirical
data = 0.95). For all 9 sites nationwide checked in this manner (namely, Ballyhaise, Belmullet, Carlow Oakpark, Casement Aerodrome, Dublin Airport, Fermoy Moorepark, Malin Head, Roches Point, and Valentia Observatory), there was an overlap of $>$ 0.89 between the overlaid densities of exceedances of the threshold for the observations and the gridded data from the nearest gridpoint.

\subsection{Temperature Anomalies}

In order to focus our research on maximum temperatures which would be considered extreme relative to the time of year in which they occurred, we chose to create and then analyse temperature anomalies. Adapting a procedure used by \citet{brown2008global}, we created the anomalies by removing the mean annual cycle at each location. To do this, we first calculated a 31-day moving average for each location over the 30 years of data, using only values at that location. We then averaged by calendar date in order to get a climate value for the first of January, a climate value for the second of January \textit{etc.} at that location. Using these 366 values provided a distinct climate curve for each location - a mean annual cycle at that gridpoint. Each of these values were then subtracted from the 30 years of raw data for the corresponding day, resulting in a dataset of anomalies at each location. It is the extreme values of these anomalies that we wish to analyse.
\\

\begin{figure}[H]
    \centering
    \begin{subfigure}{.5\textwidth}
  \centering
  \includegraphics[width=1\linewidth,keepaspectratio]{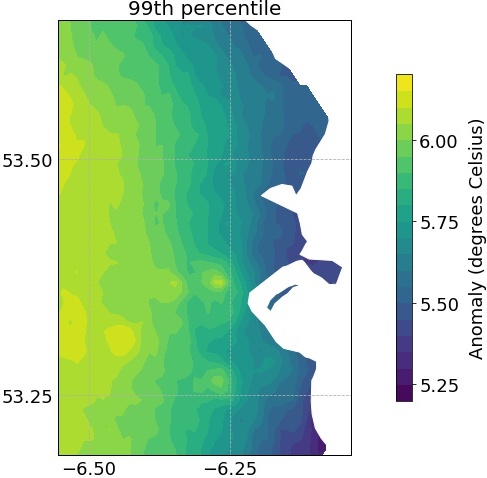}
\end{subfigure}%
\begin{subfigure}{.5\textwidth}
  \centering
  \includegraphics[width=0.92\linewidth,keepaspectratio]{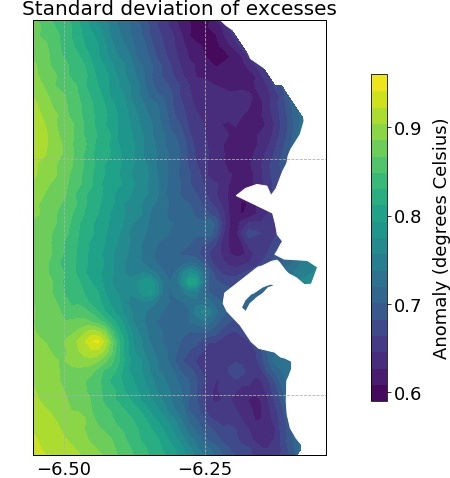}
\end{subfigure}
\caption{The figure on the left (a) is the threshold anomaly (the 99th percentile) - that is, the value for each location above which only 1\% of observations lie; the figure on the right (b) illustrates the standard deviation of the excesses of this threshold. Both scales are in degrees Celsius.}
\label{fig:anomaly}
\end{figure}

We first selected those anomalies which exceeded the 99th percentile at their gridpoint, in order to further explore the values which would be considered extreme. Figure \ref{fig:anomaly}(a) illustrates this threshold surface: it shows the value at each point above which only 1\% of anomalies at that point lie, and ranges here from 5.2 - 6.2\degree{}. The threshold is generally lower near the coast - this is unsurprising, as the temperatures here are moderated by their proximity to the sea. Examining the standard deviation of the excesses at each point (Figure \ref{fig:anomaly}(b)) shows a similar trend. Not only are values of the threshold higher as you move away from the coast - the excesses above this threshold are also more variable, indicating more extreme temperature anomalies inland.

\section{Methodology}

\subsection{Extreme Value Theory}

A comprehensive introduction to the field of extreme value theory (EVT) may be found in \citet{coles2001introduction}.
One common approach is to model the block maxima. We consider a sequence of independent and identically-distributed random 
variables, $Z_1, Z_2, \ldots$, and let $M_n=\max \left(Z_1,\ldots,Z_n\right)$ be the maximum over a block of $n$ values; for example, 
we may take $M_n$ to be the annual maxima in a multi-year set of daily maximum temperature data. The extremal types theorem states 
that, under certain regularity conditions, the distribution function of the $M_n$ will converge to a specific three-parameter distribution, known as the generalised extreme value (GEV) distribution. A major disadvantage to this approach is the fact that, by using only the maxima from a given block size, the data selected may not fully capture all extreme events \citep{brown2008global}. For example, the two most extreme events in a dataset may occur in the same year - with an annual maxima approach, only one of these will be retained.
\\

Here we consider our dataset of daily temperature maximum values, $T_{max}$ (the fact that these are anomalies is omitted from the notation, but should be remembered). Choosing to model this dataset with, for example, annual maxima would be quite inefficient, leading to posterior parameter estimates with large variance. An alternative is to model the excesses over a given threshold, often called a peaks-over-threshold (POT) approach \citep{pickands1975statistical}. For this, we assume that our sequence of independent random variables, $Z_1, Z_2, \ldots$, satisfies the extremal types theorem described above. For a large enough threshold $u$, the distribution function of the excesses $Y = Z - u$, conditional on $Z > u$, is described approximately by the generalised Pareto distribution (GPD) with cumulative distribution function:
\begin{equation}\label{eqn:gpd}
F(y) = 1 - \left( 1 + \frac{\xi y}{\sigma} \right)^{-1/\xi}
\end{equation}
defined on the set $ \left\{ y : y>0 \enskip {\rm and } \enskip (1+\xi y/\sigma)>0 \right\} $. Here, $\xi$ and $\sigma$ are known as the shape and scale parameters, respectively, and have ranges $-\infty < \xi < \infty$ and $\sigma > 0$.
\\

Both the block-maxima and the POT approach to EVT have been applied extensively to maximum temperatures from different sources. Examples of the use of GEV models for maximum temperatures are to be found in \citet{plavcova2011evaluation}, \citet{kharin2013changes}, and \citet{wang2016evaluation}. Examples of the use of GPD models for maximum temperatures can be found in \citet{laurent2007estimation}, \citet{brown2008global}, and \citet{osman2015modelling}.
In addition, a number of papers have compared the two approaches; see, for example, \citet{unkavsevic2009changes}, \citet{parey2010different}, and \citet{kioutsioukis2010statistical}.
\\

Given the parameters of the GPD distribution, we can compute the $N$-year return level. For the GPD in (\ref{eqn:gpd}), we have 
\begin{equation}\label{eqn:Pzu}
P(Z>z | Z>u) = \left( 1 + \frac{\xi (z - u)}{\sigma} \right)^{-1/\xi}
\end{equation}
Writing $\zeta_u = P(Z>u)$, we can then find the return level $z_m$, the level which is exceeded on average once every $m$ observations, by solving:
$$
 P(Z>z_m) = \zeta_u  \, \left( 1 + \frac{\xi (z_m - u)}{\sigma} \right)^{-1/\xi} = \frac{1}{m}. 
$$
Letting $m = N\,n_y$, where $n_y$ is the number of observations per year, we arrive at the following expression for the $N$-year return level:
 \begin{equation}\label{eqn:rl_gpd}
 z_N = u + \frac{\sigma}{\xi} \left[ (N n_y \zeta_u)^{\xi}  -1 \right]
 \end{equation}

\subsection{Spatial methods}

Typically when dealing with random variables recorded as point-referenced (or geostatistical) data, the location index \textbf{s} is assumed to vary continuously over \textit{D}, a fixed subset of $\mathbb{R}^d$ \citep{banerjee2014hierarchical}. Let $T(\textbf{s})$ be a vector of random variables (a random vector) at locations \textbf{s}. This could be, for example, measurements of daily maximum temperatures at locations \textbf{s}. While it is sensible to conceptually assume such values exist at all possible sites in the spatial domain, in practice the data is a partial realisation of this continuous spatial process. Given this partial realisation, the problem then becomes inference about this spatial process $T(\textbf{s})$ as well as prediction at new locations. To this end, it is assumed that the covariance between the random variables at two locations depends on the distance between these locations. That is, $Cov(T(s_i), T(s_j)) = C(s_i, s_j) = C(d_{ij})$ is a function of $d_{ij}$ where $d_{ij}$ is the distance between locations $s_i$ and $s_j$ (for brevity, the dependence on sites $i$ and $j$ is dropped below). The method of calculating this distance must be specified (with Euclidean distance the most common approach). There are many choices of covariance functions (see \citet{gelfand2010handbook} for a description of several parametric models, and their relative merits). In this work, we use the Mat\'ern class of covariance functions \citep{matern2013spatial} with univariate form:

$$
C(d) = \begin{cases}
\frac{\varsigma^2}{2^{\nu - 1} \Gamma(\nu)} (\phi d)^{\nu} K_{\nu} (\phi d) & \hspace{3 pt} \text{if} \hspace{3 pt} d \neq 0 \\
\tau^2 + \varsigma^2 & \hspace{3 pt} \text{if} \hspace{3 pt} d = 0.
\end{cases}
$$

Here, $\varsigma^2$ is the partial sill (variance of the spatial effect), $\tau^2$ is the nugget (variance of the non-spatial effect), $\nu$ is a parameter controlling the smoothness of the spatial field, $\phi$ is a spatial decay parameter controlling how quickly the covariance decreases with distance, $\Gamma()$ is the gamma function, and $K_{\nu}$ is the modified Bessel function of the second kind of order $\nu$.
\\

There are alternatives to using the Mat\'ern class of covariance functions, such as kernel convolution (or moving average) models, and convolutions of covariance models \citep{gelfand2004nonstationary}. However, we decided to work with the Mat\'ern class of functions as it is a flexible class, with parameters that have attractive interpretations, and includes as special cases the exponential and Gaussian covariance functions \citep{banerjee2014hierarchical}.

\subsubsection{Gaussian Processes} \label{GPs}

The process $T(\textbf{s})$ is said to be Gaussian if, for any $n \geq 1$ and any set of sites $\{s_1, s_2, ... s_n\}$, $\textbf{T} = (T(s_1), T(s_2), ... T(s_n))^T$ has a multivariate normal distribution \citep{banerjee2014hierarchical}. Gaussian Processes (GPs) can be thought of as extending the finite multivariate normal distribution to infinitely many random variables; in other words, a GP is an infinite collection of variables such that every finite subset follows a multivariate normal distribution. This is a very flexible framework for modelling spatial data, as the covariance matrix can be specified using any valid covariance function.
\\

Given realisations of the the process $T(\textbf{s})$, and $p$ spatially-referenced covariates at the same locations \textbf{s}, let $X(\textbf{s})$ be the $n \times (p+1)$ matrix associated with the spatial regression model:

$$
T(\textbf{s}) = X^T(\textbf{s}) \alpha + w(\textbf{s}) + \epsilon(\textbf{s})
$$

where $X^T(\textbf{s}) \alpha$ is the mean response, $w(\textbf{s})$ is a zero-centred GP with covariance function $ C(\textbf{s}, \textbf{s}')$ and $\epsilon(\textbf{s}) \overset{iid}{\sim} N(0, \tau^2)$ is an independent measurement error (see, e.g., \citet{finley2009improving} and \citet{cressie2015statistics}).
\\

There are alternative spatial methods to GPs such as the integrated nested Laplace approximation (INLA) approach proposed by \citet{rue2009approximate}, and INLA combined with a stochastic partial differential equation approach (INLA-SPDE) proposed by \citet{lindgren2011explicit}. However, we have chosen to use GPs due to the ease with which they fit into a Bayesian hierarchical framework. They are flexible empirical models, which are appropriate for an irregularly fluctuating and real-valued spatial surface \citep{diggle2007springer}, as we have here.
\\

Bayesian inference using GPs typically involves the need to invert or factor the covariance matrix. This becomes computationally impractical as the dimension $n$ becomes large (that is, a large number of gridpoints), particularly when using an algorithm such as Markov chain Monte Carlo (MCMC) which involves inverting or factoring this matrix hundreds of thousands of times in one run of a model fitting. For this reason, we decided to use reduced-rank representations focusing on Gaussian predictive process models.

\subsubsection{Gaussian Predictive Process Models} \label{GPsection}

A comprehensive overview of hierarchical Gaussian predictive process models is given in \citet{banerjee2014hierarchical}. We offer a brief summary below. Following the notation from section \ref{GPs}, we can avoid dealing with the dense covariance matrix induced by the zero-centred GP $w(\textbf{s})$ by projecting it onto a subspace spanned by its realisation over the $n^*$-dimensional $S^*$ where $n^*$ $ \ll$  $n$. An optimal projection of the process $w$ at location $s$, based upon its realisation over $S^*$ is given by the kriging equation:

\begin{equation}\label{eqn:kriging}
\tilde{w}(s) = E(w(s) | w^*)
\end{equation}

where $w^* = (w(s_1^*), w(s_2^*), ... w(s_n^*))^T$. $\tilde{w}(s)$ is referred to as the predictive process derived from the parent process $w(s)$.
\\

Further expanding the above:

$$
\tilde{w}(s) = C(s, S^*)^T \hspace{5 pt} C(S^*, S^*)^{-1} \hspace{5 pt} w(S^*)
$$

where $C(s, S^*)$ is the $n^* \times 1$ vector with $C(s, s_j^*)$ as the $j^{th}$ element; $C(S^*, S^*)^{-1}$ is the $n^* \times n^*$ covariance matrix on the subspace $S^*$; and $w(S^*)$ is the $n^* \times 1$ values of the process $w$ on the subspace $S^*$. The important point now is that inference on the process $w$ will now involve inverting or factoring an $n^*$-dimensional matrix rather than an $n$-dimensional one.

\subsection{Hierarchical Model}

The aim of this study is to produce $N-$year return levels of anomalies of extreme temperature ($T_{max}$). The dataset we want to model has already been described. In summary, anomalies were calculated at each location; site-specific thresholds ($u_i$) of the 99.5th percentile were calculated; only declustered excesses above this threshold were kept for modelling. We require location-specific estimates of the parameters from the GPD ($\sigma_i$ and $\xi_i$), as well as the probability of exceeding the threshold ($\zeta_i$) in order to calculate return levels using equation (\ref{eqn:rl_gpd}).
\\

\begin{figure}[h]
    \centering
       \includegraphics[width=1\textwidth,keepaspectratio]{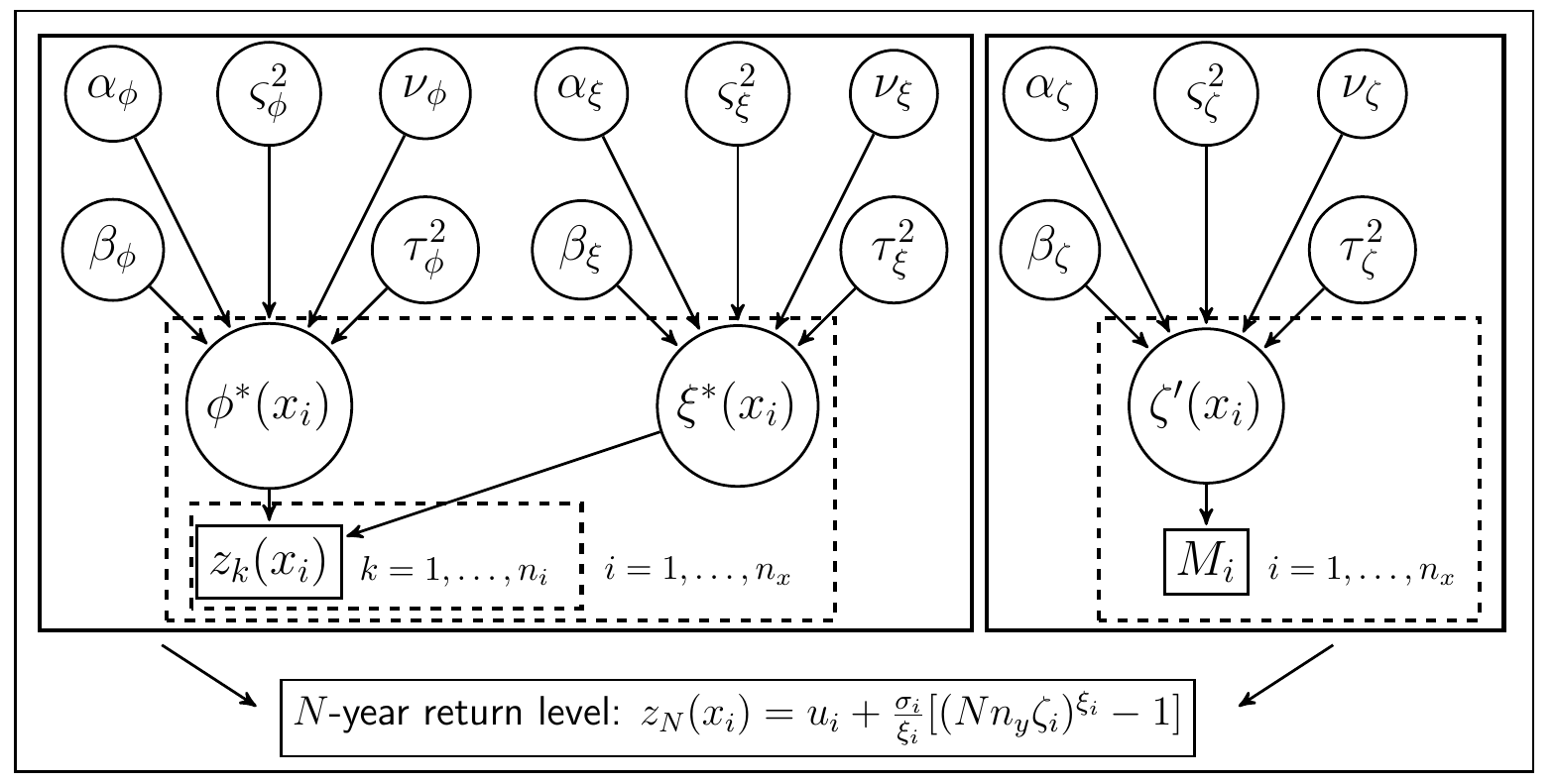}
      \caption{A directed acyclic graph (DAG) of the Bayesian hierarchical model fitted to the spatial dataset. Details of each layer and the parameters involved may be found in the text.}
      \label{fig:DAG}
\end{figure}

Following the approach of \citet{cooley2007bayesian}, we employ a Bayesian hierarchical model with three layers (see the directed acyclic graph in Figure \ref{fig:DAG}). There are two separate hierarchies - both with three layers. The first hierarchy models the parameters of the GPD; the second models the probability of exceeding the threshold. The first layers in both of these hierarchies consist of modelling the data; the second layers describe the latent spatial process underlying the extremes; while the third layers consist of the prior distributions on the parameters controlling the second.

Let the declustered extreme anomalies of $T_{max}$ be denoted by $z_k(x_i)$ where the indices $i$ and $k$ are such that $z_k(x_i)$ refers to the $k$-th exceedance ($k = 1 \dots n_i$) at gridpoint $x_i$ ($i = 1 \dots n_x$). Let $M_i$ refer to the number of exceedances. The layers in the hierarchies are described in detail below.

\subsubsection{Layer 1: Extremal Data and Probability of Exceedance}

We model the extremal data $z_k(x_i)$ using the GPD. This is the first layer in the first hierarchy. To ensure a positive scale parameter throughout the computations, we reparameterise $\phi = log(\sigma)$. We have two spatially-varying parameters for the distribution. The first layer in the first hierarchy is then given by:

$$
z_k(x_i) \sim GPD(u(x_i), \sigma(x_i), \xi(x_i))
$$

The first layer in the second hierarchy involves the parameter $\zeta_i$ (the probability of exceeding the threshold $u_i$ - or more precisely, following declustering, the probability of being a cluster maximum at gridpoint $i$). This needs to be modelled in order to calculate return level surfaces using equation (\ref{eqn:rl_gpd}). Following the methodology of \citet{cooley2007bayesian}, we model this as a binomial random variable. Here, the probability of being a cluster maximum is modelled using the empirical probability as our data. It is assumed that the observed number of cluster maxima $M_i$ at gridpoint $i$ is a binomial random variable with $m_i$ trials (the total number of observations in the period of study), each with a probability $\zeta(x_i)$ of being a cluster maximum:

$$
M_i \sim Bin(m_i, \zeta(x_i))
$$

From this point, we omit all dependence on location $x_i$ for ease of notation, unless emphasis at a particular point is necessary.

\subsubsection{Layer 2: Process}

We assume that the GPD parameters vary smoothly over space and thus model the two variables ($\phi=log(\sigma)$ and $\xi$) as GPs. Following our decision to model these parent processes using the reduced-rank representation of predictive processes, at this layer we thus directly model $\phi^* =log(\sigma^*)$ and $\xi^*$ (where $\phi^*$ and $\xi^*$ are defined on a smaller grid, and are thus lower-dimensional, than $\phi$ and $\xi$, which are defined on the full grid). The second layer for $\phi^*$ in this hierarchy then is:

$$
\phi^* \sim MVN(\mu_\phi, \Sigma_\phi)
$$

Here, $\mu_\phi = X \alpha_\phi$ where $X$ is a matrix of covariates for the linear regression component of the model, and $\alpha_\phi$ is a vector of coefficients. $\Sigma_\phi$ is the covariance matrix of the spatial process, and is modelled using the Mat\'ern covariance function described earlier. This has parameters $\beta_\phi$ consisting of a matrix of range parameters (controlling how quickly the correlation drops off in different directions), $\varsigma_\phi$ is the partial sill, $\tau_\phi$ is the nuggest, and $\nu_\phi$ is the smoothness parameter. The $n$-dimensional $\tilde{\phi}$ is then calculated using the kriging equation (\ref{eqn:kriging}) detailed in section \ref{GPs}. The layer for $\xi^*$ is similar.
\\

In a similar manner (and again following the methodology of \citet{cooley2007bayesian}), we assume $\zeta$ to vary smoothly over space. We let $\zeta^*$ be a reduced-rank representation of $\zeta$, and then apply the transformation $\zeta' = \text{logit}(\zeta^*)$, and use predictive processes to model this reduced-rank transformed representation. As before, the kriging equation is used to calculate $\tilde{\zeta}$ given $\zeta'$.

\subsubsection{Layer 3: Hyperparameters}

The third and final layers of the hierarchies consist of the 15 prior distributions on the parameters in the second layers - that is, the distributions of the $\alpha$, $\beta$, $\varsigma$, $\tau$ and $\nu$ hyperparameters, for each of the three parameters $\phi^*$, $\xi^*$ and $\zeta'$.
\\

Using Bayesian inference allows additional information about a process to be incorporated in the form of prior information. This could be of great benefit in our case, due to the scarcity of extreme data. However, \citet{coles1996bayesian} argue that with such scarce data, an expert may not be able to independently formulate prior beliefs about this process. With this in mind, we aim to use semi-informative prior distributions which are based on physically plausible values, but with enough flexibility so that the data is not restricted from informing the posterior distributions.
\\

We constructed the regression matrix $X$ to model an intercept parameter in addition to the covariates of latitude, longitude, and elevation. As elevation had a strong positive skew (skewness = 2.36), we transformed it using a log-transform. The three covariates were then scaled to be centred on 0 and with a standard deviation of 1.
\\

For the prior distributions for the corresponding vectors of regression coefficients $\alpha_{\phi}$, $\alpha_{\xi}$, and $\alpha_{\zeta}$, we chose Normal distributions centred on 0, 0 and -6 respectively for the intercept parameters. Remembering that $\phi^*$ is the log of the scale parameter, this corresponds to a prior distribution for the scale parameter $\sigma$ which is centred on 1. The prior for the intercept for $\xi^*$ is centred on 0, which assumes the data to have infinite support. The prior for the intercept for $\zeta'$ is centred on -6, which is approximately the logit of the probability of an observation selected at random exceeding the chosen threshold. Standard deviations of 2 are used for all three intercepts. This is arguably a little too wide in some cases (e.g., \citet{coles1996bayesian} point out that a shape of -1 rarely occurs when modelling the maxima of environmental data), but as the shape parameter is particularly difficult to model, we preferred to err on the side of caution, and allow the data play the dominant role in informing the posterior distribution for this parameter.
\\

The following details regarding the non-intercept $\alpha$ coefficients, and the $\beta$, $\nu$, $\varsigma^2$ and $\tau^2$ hyperparameters are identical for the $\phi^*$, $\xi^*$ and $\zeta'$ parameter surfaces. For the non-intercept $\alpha$ coefficients for the scaled covariates of latitude, longitude and altitude, we chose $N(0, 1)$ distributions (as proposed by \citet{dyrrdal2015bayesian}). Although we expect to see a relationship between the covariates and the parameter surfaces (e.g, more extreme excesses are expected further from the sea, and so the scale surface is likely to increase as the distance from the sea increases), we wanted to assume no relationship \textit{a priori} and allow the data to inform the relationship between the parameter surfaces and the covariates. A standard deviation of 1 is, again, arguably a little too wide (with the scaled covariates, a value of 1 would imply a parameter surface which increases by at least 3 across the fields of latitude, longitude or altitude - a slope of this magnitude is unrealistic for all three parameters under consideration in this study). We have aimed to use physical arguments in informing the prior distributions, but again have chosen to err on the side of caution in order to allow the data play the dominant role in informing the posterior distributions. 
\\

The $\beta$ and $\nu$ priors needed to be considered together. Looking again at the univariate Mat\'ern function for the covariance between two gridpoints, where $d$ is the distance between the gridpoints $s$ and $s'$:

$$
C(s, s') = C(d) = \begin{cases}
\frac{\varsigma^2}{2^{\nu - 1} \Gamma(\nu)} (\beta d)^{\nu} K_{\nu} (\beta d) & \hspace{3 pt} \text{if} \hspace{3 pt} d \neq 0 \\
\varsigma^2 + \tau^2 & \hspace{3 pt} \text{if} \hspace{3 pt} d = 0,
\end{cases}
$$

it can be seen that the shape of the parameter surface depends on the spatial decay parameter $\beta$ (which controls how quickly the covariance decreases with distance) and $\nu$ (which controls the smoothness of the spatial field).
\\

In order to allow the spatial decay parameter $\beta$ to differ depending on the direction of its two-dimensional coordinates (longitude and latitude), it is necessary to extend the univariate case above to incorporate a $2\times2$ positive definite matrix $\beta$. With this approach, the (1, 1) entry in the matrix represents the spatial decay in the direction of longitude, the (2, 2) entry represents the spatial decay in the direction of latitude, and the (1, 2) entry represents the covariance between the two. The covariance function between two points $s$ and $s'$ now becomes:

$$
C(s, s') = C(d) = \begin{cases}
\frac{\varsigma^2}{2^{\nu - 1} \Gamma(\nu)} \Big(\sqrt{d^T \beta^{-1} d}\Big)^{\nu} K_{\nu} \Big(\sqrt{d^T \beta^{-1} d}\Big) & \hspace{3 pt} \text{if} \hspace{3 pt} d \neq 0 \\
\varsigma^2 + \tau^2 & \hspace{3 pt} \text{if} \hspace{3 pt} d = 0,
\end{cases}
$$

where $d$ is now a 2-dimensional vector of the distance between the two points.
\\

In order to construct a set of prior matrices $\beta$, all possible combinations of matrices were formed using values from the set \{0, 0.05, 1, 10\}, from which only the positive definite ones were retained. This led to a set of 20 matrices. The smoothness parameter $\nu$ is assigned a prior support of \{0.5, 2.5\}. It is common to use such values for $\nu$, as the data can rarely inform about smoothness of higher orders \citep{finley2009improving}. This means there are 20 x 2 = 40 prior ($\beta$, $\nu$) combinations to model the spatial decay across the parameter field and the smoothness of the resulting field.
\\

Though it is possible for the spatial dependence of individual extreme observations to have a short range, the surfaces being modelled here refer to a climatological quantity rather than a weather quantity - and so it is reasonable to assume the climate will be similar at two nearby locations \citep{cooley2007bayesian}. With this in mind, the ($\beta$, $\nu$) combinations above allow for either very short or very long effective spatial ranges, which we take to be the distance at which the correlation equals 0.05 \citep{finley2013spbayes}. The extent of these combinations means that correlation can either drop off very quickly with distance (with an effective spatial range of 10 km) or else reduce very slowly (reducing from 1 to 0.67 between the two furthest points on the grid). This latter combination means all points can be well within the effective spatial range of all other points in the domain of the study. We feel that this represents a sufficiently broad selection to allow for great flexibility in modelling the spatial fields of the parameters. Though physically plausible (but conservative approach), we will check that the posteriors for $\beta$ assign negligible probabilities to the extreme combinations in the specified prior range \citep{diggle2007springer}.
\\

The remaining two parameters in the Mat\'ern covariance function, $\varsigma^2$ and $\tau^2$, represent the partial sill (variance of the spatial effect) and the nugget (variance of the non-spatial effect - essentially representing the measurement error in repeated measurements at any site) respectively. It is difficult to have information on these parameters \textit{a priori}, so we chose relatively uninformative priors. As both of these quantites are positive, we chose to model their log: $\text{log}(\varsigma^2) \sim N(0, 1)$ and $\text{log}(\tau^2) \sim N(-2.3, 1)$ (i.e., we assume $\varsigma^2$ has a mode of 1, and $\tau^2$ has a mode of $\sim 0.1$). The standard deviations of 1 (for the priors on $\text{log}(\varsigma^2)$ and $\text{log}(\tau^2)$) are again sufficiently large to allow the data play the dominant role in informing the posterior distributions for these parameters, without allowing for unrealistic larger values to occur.

\subsection{Threshold selection}

In order to ensure independence of observations in time, we declustered the dataset of extremes anomalies by removing all observations which occurred in clusters except for the maximum of this cluster (a process described by \citet{coles2001introduction}, p.99). That is, if two or more consecutive days at any point exceeded the threshold, only the maximum of these values was retained. The model was run for varying thresholds from the 98th percentile upwards. As detailed in \citet{coles2001introduction}, threshold selection involves a trade-off between bias and variance (higher thresholds lead to reduced bias, but increased variance). In our case, choosing too low a threshold will result in many less-extreme exceedances (i.e., extremes which are marginally above the threshold), which will threaten the asymptotic nature of the GPD model. Choosing too high a threshold will result in too few datapoints, and result in large uncertainties in the posterior estimates. Here, we selected a threshold of the 99.5th percentile - results for simulations below this tended to be overly dominated by the large number of excesses marginally above the threshold, and failed to model the extreme excesses sufficiently well. Following declustering of excesses above this 99.5th percentile threshold, there was a median number of 40 independent excesses retained across the domain (the 2.5th percentile value of the number of independent excesses retained was 35, and the corresponding number for the 97.5th percentile value was 50 (see Table \ref{tab:table1})).

\begin{table}[H]
  \begin{center}
       \begin{tabular}{c|c|c|c|c|c|c}
      \toprule 
      \textbf{\diagbox{Percentile}{Threshold}} & 98th & 98.5th & 99th & 99.5th & 99.7th & 99.9th \\
      \midrule 
	 2.5\% &  136 &  105 &  73 &  35 &  21 & 8 \\
	 25\% &   148 &  114 &  76 &  37 &  22 & 9 \\
      50\% &   152 &  117 &  79 &  40 &  27 & 10 \\
	 75\% &   162 &  126 &  86 &  46 & 29 & 11 \\
	 97.5\% & 176 & 137 &  93 &  50 & 31 & 11 \\
      \bottomrule 
    \end{tabular}
  \end{center}
      \caption{This table shows, for six selected threshold surfaces of the 98th percentile and above, the remaining number of declustered excesses across the surface for the 2.5\%, 25\%, 50\%, 75\% and 97.5\% percentile levels. For example, the 98th percentile threshold has a median number of excesses at a gridpoint of 152, with a 2.5th percentile value of 136 excesses and a 97.5th percentile value of 176 excesses.}
    \label{tab:table1}
\end{table}

\subsection{Model implementation}

We implemented our model using the programming language R \citep{R} and the package Rcpp \citep{Rcpp}. A Metropolis-Hastings MCMC algorithm was used to draw samples from the posterior distributions of all parameters in the hierarchy. For both components of the model, three chains were run for 50,000 iterations. A burn-in of 10,000 iterations was discarded from each chain. The remaining chains were then thinned by retaining only every 10th sample to reduce auto-correlation. Convergence was then assessed using the $\hat{R}$ criterion \citep{gelman1992inference}, with values below the suggested criterion of 1.2 taken to imply convergence. The resulting simulations are presented in the next section. All code needed to reproduce this analysis is available in a public repository on GitHub at \\
\href{https://github.com/jackos13/extremes}{https://github.com/jackos13/extremes}.


\section{Results}

\subsection{Posterior parameter estimates}

\begin{figure}[H]
    \centering
    \begin{subfigure}{.5\textwidth}
  \centering
  \includegraphics[width=1\linewidth,keepaspectratio]{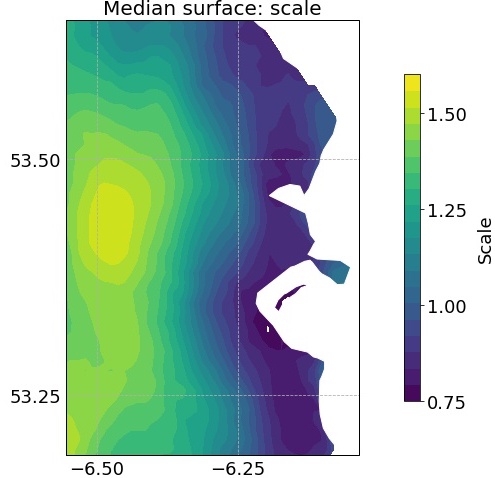}
\end{subfigure}%
\begin{subfigure}{.5\textwidth}
  \centering
  \includegraphics[width=0.92\linewidth,keepaspectratio]{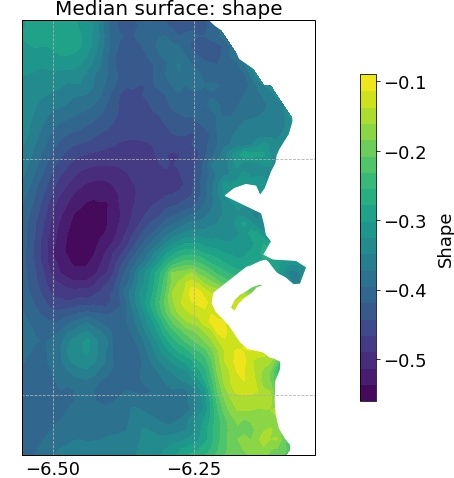}
\end{subfigure}
\caption{Shown are the posterior estimates for the median surface of the scale parameter on the left (a) and the shape parameter on the right (b).}
\label{fig:params}
\end{figure}

Posterior median surfaces for the scale and shape parameters of the GPD are shown in Figure \ref{fig:params}. The scale parameter (indicating the variance of the distribution) increases with increasing distance from the coast. Along the coastline, it has a median value of between 0.75 and 1. Further inland, it exceeds 1 at most locations, extending upwards of 1.5 in the west and south-west of the study domain. This is consistent with the surface of the standard deviation in Figure \ref{fig:anomaly}(b), which showed that the excesses of the threshold farther from the coast had larger variance. In other words, more variable temperature extremes are observed farther from the coast, due to the increased distance from the moderating effect of the Irish Sea. This corresponds to standard meteorological theory where the diurnal range of temperature generally increases with distance from the sea \citep{rafferty2011climate}.

The shape parameter is slightly more difficult to interpret. A negative shape indicates a finite upper-bound to the corresponding posterior distribution; a shape of 0 indicates that the data has infinite support; while a positive shape indicates a finite lower-bound to the corresponding posterior distribution. The scale and shape are known to be generally negatively correlated \citep{cooley2010spatial}. This is evident here, where the median shape is seen to have the opposite trend to the median scale - the shape surface generally decreases with increasing distance from the coast (Figure \ref{fig:params}(b)).
\\

\begin{figure}[H]
    \centering
  \includegraphics[width=0.5\linewidth]{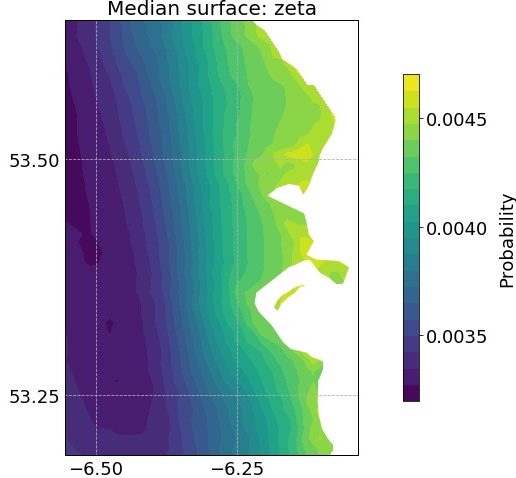}
\caption{Shown is the posterior estimate for the median surface of the zeta parameter, the probability of a randomly selected day being a cluster maximum.}
\label{fig:params2}
\end{figure}

The median surface of the zeta parameter is shown in Figure \ref{fig:params2}. The relatively simple nature of this parameter (essentially, it models the binomial probability that a randomly chosen day is a cluster maximum) led it to converge quite quickly, with very little uncertainty in its posterior distribution. The general pattern here is for higher values nearer the coast. This indicates that extreme temperatures exceeding the threshold are more likely to be isolated incidents here, whereas further inland extremes are more likely to occur in clusters (and therefore there are fewer independent excesses retained here). The higher frequency of isolated excesses near the coast is due to the moderating effect of the sea, which makes it more unlikely for prolonged periods where successive days exceed the threshold to occur.

\subsection{Posterior return level estimates}

\begin{figure}[H]
    \centering
  \includegraphics[width=0.8\linewidth,keepaspectratio]{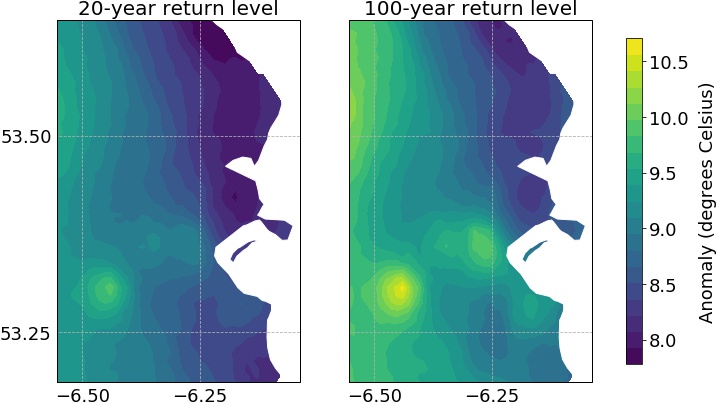}
\caption{Shown are the 20- and 100-year return level median surfaces of the temperature anomaly.}
\label{fig:rlsurfaces}
\end{figure}

20- and 100-year return-level median surfaces are shown in Figure \ref{fig:rlsurfaces}. The 20-year median surface is seen to range from just below 8\degree{C} to almost 10\degree{C}. The lowest levels are seen along the north-eastern coast of the study domain, while higher return levels are observed inland. The highest return levels are observed on the western side of the domain. The 100-year median surface ranges from 8\degree{C} to just under 10.7\degree{C}. A similar pattern to the 20-year return level is observed - lowest values appear along the sea in the north-east of the domain, with highest values on the western boundary. The highest part of the mean climate curve calculated for Casement Aerodrome (its location is shown in Figure \ref{fig:dublin}(b)) exceeds  20\degree{C} in July. This means that if the temperature anomaly return levels seen in Figure \ref{fig:rlsurfaces} occur at this time of year, daily maximum temperatures will be in excess of 30.5\degree{C}. Since observations began at this location in 1944, there have been only two days recorded here with temperatures exceeding 30\degree{C} - one in 1975 (before the time period of this study) and one in 2006.

\subsection{Comparison with recent observations}

\begin{figure}[H]
    \centering
        \includegraphics[width=1\textwidth,keepaspectratio]{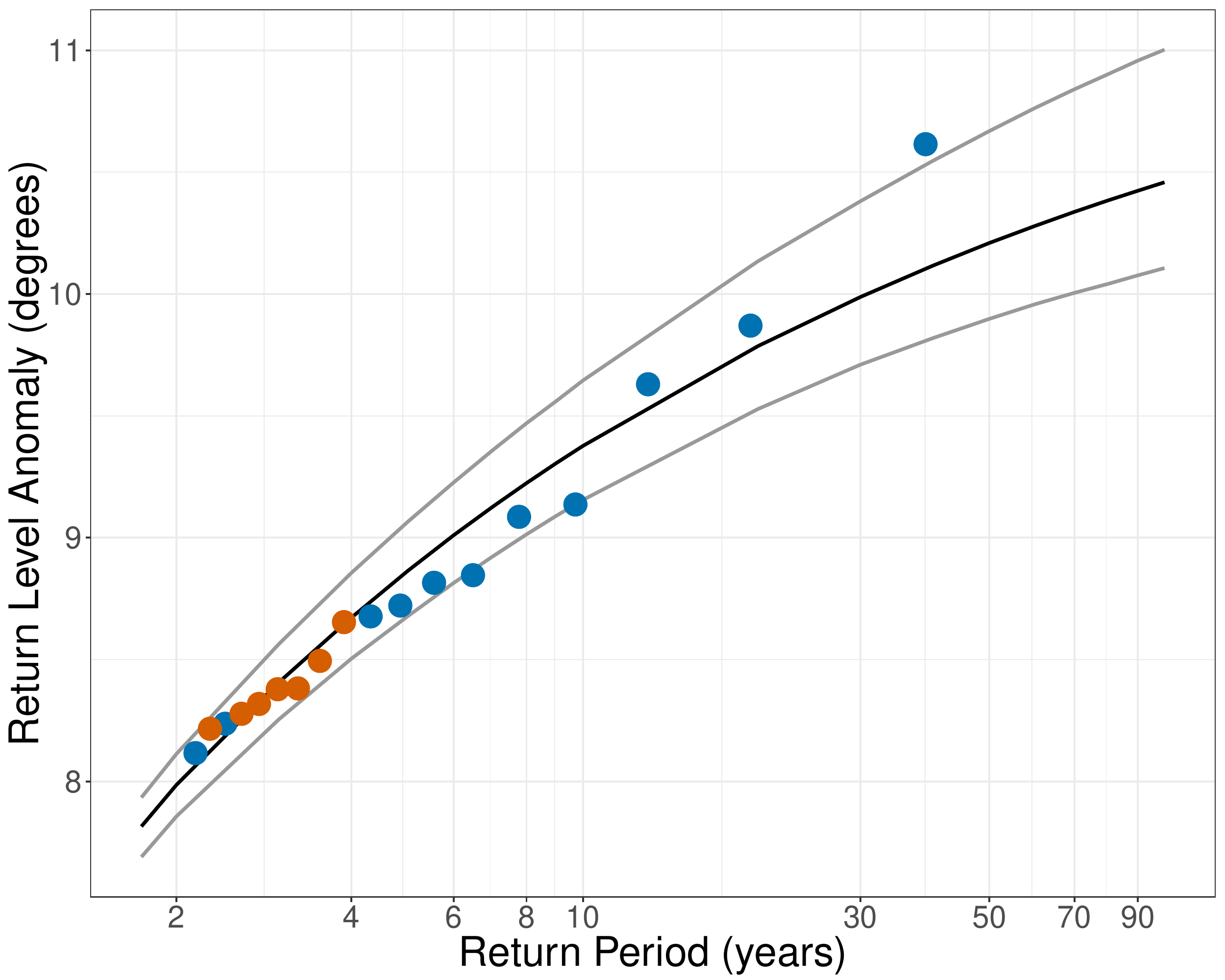} 
    \caption{Shown is the return level curve for Casement Aerodrome showing the median (black) and the 95\% credible interval (grey) curves, with observations superimposed for 1981-2010 (blue dots) and 2011-2018 (orange dots).}
    \label{fig:rlcurves}
\end{figure}

Figure \ref{fig:rlcurves} shows the return level curve at Casement Aerodrome (location shown in Figure \ref{fig:dublin}), with the observations from the synoptic station overlaid on the plot. The return period (measured in years) is displayed on a log scale for ease of interpretation. The median return level curve is shown in black; a 95\% credible interval is contained within the upper and lower bounds in grey. Observations are included by plotting their empirical return period against their return level anomaly. This is found by ordering the observed excesses in a vector $y$ such that $y_1 \leq y_2 \leq \dots \leq y_n$ for the $i = 1 \dots n$ excesses at that location, and then calculating the corresponding vector of return periods $x$ with entries given by:

$$
x_i =  \frac{1}{1 - \sfrac{i}{(n+1)}} \times \frac{1}{npy}
$$

Here $npy$ refers to the number of observations per year. Observations from the time period of the study (1981-2010) are plotted with blue circles. Observations from the most recent eight years (2011-2018) are plotted with orange circles. There were more excesses per year in the period 2011-2018 (12 in total - 1.5 per year) than there were in the period of the study (34 in total - 1.33 per year). However, this increase in the frequency of threshold excesses does not appear to be due to an increase in the severity of threshold excesses: as can be seen in the plot, all of the more recent excesses appear in the lower region of the return level curve. For reasons of clarity, the graph only includes those points with a return period of 2 years or greater. 11 points from 1981-2010 are included in this set, and 7 points from 2011-2018. The greatest empirical return period calculated for an anomaly from 1981-2010 is almost 39 years, whereas the greatest empirical return period calculated for an anomaly for the 2011-2018 period is just under 4 years. This again demonstrates the increase in the frequency of threshold excesses at this location, but shows that there is no corresponding increase in their severity.

\begin{figure}[H]
    \centering
        \includegraphics[width=1\textwidth,keepaspectratio]{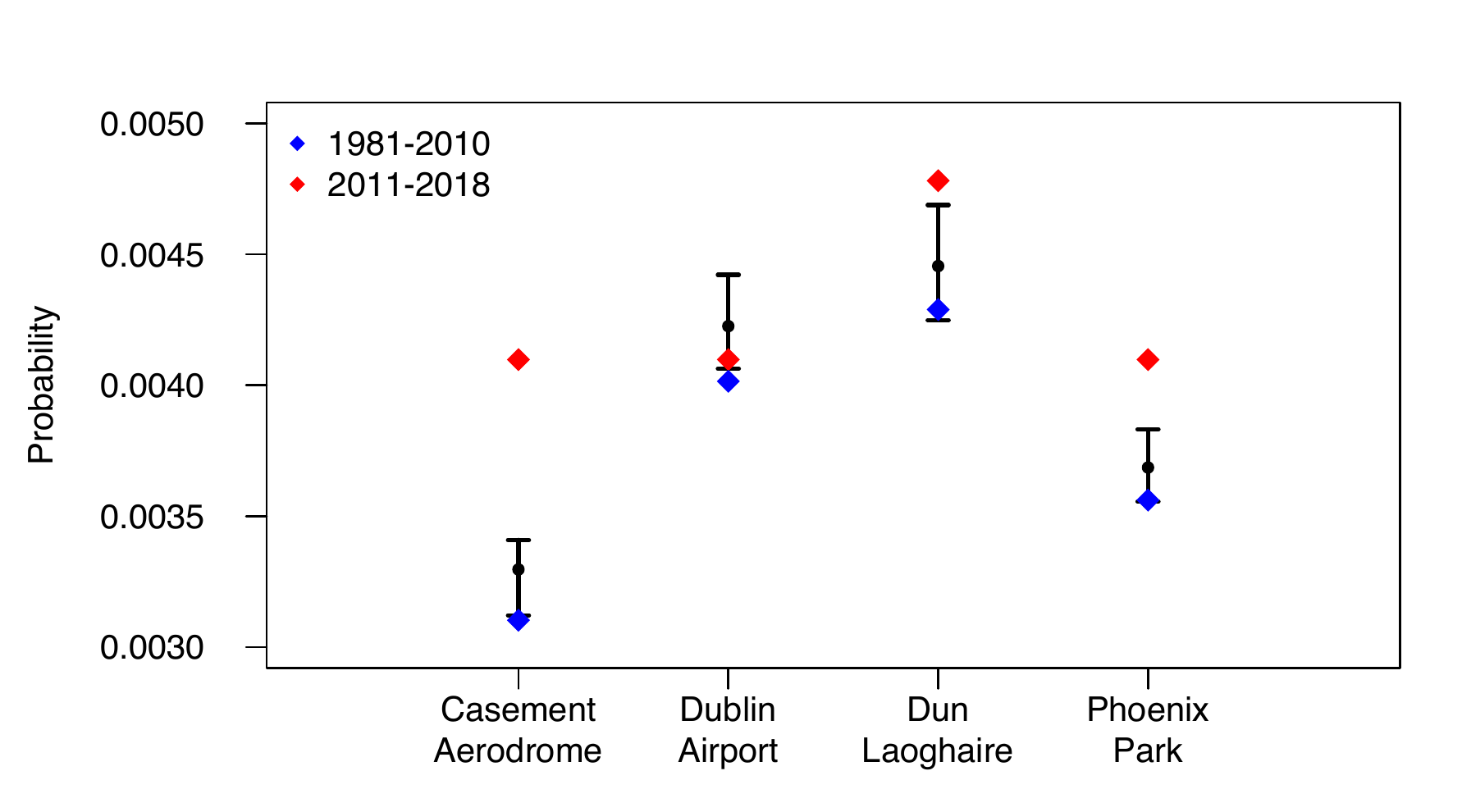} 
    \caption{Shown in black is the 95\% credible interval of the probability of being a cluster maximum ($\zeta$) at four synoptic stations: Casement Aerodrome, Dublin Airport, Dun Laoghaire, and the Phoenix Park; black dots show the posterior median values; blue diamonds show the observed (1981-2010) probability of being a cluster maximum - that is, the data used to fit the model; red diamonds show the more recent (2011-2018) observed probability of being a cluster maximum.}
    \label{fig:newzeta}
\end{figure}

This increase in the frequency of threshold excesses from recent observational data is seen at more than one location. Figure \ref{fig:newzeta} shows posterior 95\% credible intervals (black line segments, with a filled circle indicating the median) of the probability of a particular day being a cluster maximum at four synoptic stations across the domain of the study: Casement Aerodrome, Dublin Airport, Dun Laoghaire, and the Phoenix Park. Superimposed on these credible intervals is the data used to fit the model - the observed site-specific probability of being a cluster maximum for the period 1981-2010 (blue diamonds). Also superimposed is the more recent (2011-2018) observed site-specific probability of being a cluster maximum (red diamonds). From this plot, it can be seen that the greatest increase in the frequency of threshold excesses is at Casement Aerodrome (where the 2011-2018 observed probability is 20\% greater than the upperbound of the credible interval at that site). The next largest increase is seen at the Phoenix Park station, where the 2011-2018 observed probability is 7\% greater than the upperbound of the corresponding credible interval. Of the remaining two stations, the 2011-2018 observed probability at Dun Laoghaire is marginally outside the credible interval bounds (it is 2\% greater than the upperbound at this location), while the corresponding probability for Dublin Airport is within the credible interval bounds.

\section{Discussion and conclusion}

In this research, we began a comprehensive characterisation of temperature extremes in Ireland for the period 1981-2010. We produced return-level surfaces of daily maximum temperature across County Dublin, the domain of the study. We also produced site-specific return-level curves at synoptic stations, and super-imposed data from 2011-2018 onto these plots. To our knowledge, this is the first study to combine predictive processes and EVT in the manner we have used here.
\\

We modelled a spatial dataset of daily maximum temperatures over the domain of County Dublin, Ireland, in order to better understand the nature of temperature extremes there. We first created a dataset of anomalies to focus attention on temperatures which would be considered extreme relative to the time of year in which they occurred. In order to make the best use of this dataset, we chose a peaks-over-threshold approach, and declustered the exceedances of this threshold to remove successive days of extremes, retaining only the maximum excess from a cluster. We then used the GPD and the reduced-rank representation of predictive processes in the first component of two three-level Bayesian hierarchical models. This resulted in (samples of) posterior densities for the scale and shape surfaces, the parameters which uniquely determine the distribution and behaviour of the temperature anomaly excesses. The second Bayesian hierarchical model was applied in a similar manner to the surface of the probability that a day selected at random is a cluster maximum which exceeds the threshold. Our chosen Bayesian approach means that uncertainty is accounted for directly in the estimate of the full posterior density, in contrast to approaches which yield only summary statistics from the target density. Values from these posterior densities for the three surfaces were drawn at random in order to calculate estimates of return-level surfaces using equation (\ref{eqn:rl_gpd}). This direct accounting for uncertainty in the parameter distributions means that the uncertainty of the return-level surfaces can be quantified directly, as the process yields posterior densities for these values too. We used the reduced-rank representation of predictive processes to significantly reduce the computational burden of MCMC by modelling the surfaces of interest directly on a sub-grid of the domain, while still being able to incorporate and make use of the data at every gridpoint in the domain.
\\

Extremes are, by defintion, rare - our approach ensured that valuable information about the observed extremes was not neglected in pursuit of a relatively fast model-fitting algorithm. Modelling the GPD and binomial parameters as GPs which vary continuously over space also allowed us to make the best use of the limited data which we had - parameter values at a gridpoint are not only informed by the data at that gridpoint, but by the data at surrounding gridpoints too, with nearby gridpoints having more influence than those at a greater distance. This reduces the (large) uncertainties which result in parameter estimates from a single-site analysis (e.g, maximum-likelihood estimation at a single point). This is particularly helpful when it comes to the shape parameter, on which it can be very difficult to perform inference.
\\

One of the problems with modelling large spatial datasets with a likelihood-based approach using MCMC algorithms is the need to invert large and dense matrices thousands of times. We tackled this problem with the use of reduced-rank representations of the latent spatial processes, and hence the computational burden was vastly reduced. The problem of modelling extremes (that is, the unavoidable scarcity of data) was approached by using spatial models to make best use of the information contained in the data, and a Bayesian hierarchy in order to set prior distributions for the parameters which were based on physical principles.
\\

Following model fitting and convergence diagnostics, posterior parameter estimates and return level surfaces were produced. These were presented in the previous section. The median return level surfaces showed that, for example, for both 20- and 100-year time periods, exceedances of these values in July would mean maximum daily temperatures in excess of 30.5\degree{C} at Casement Aerodrome.  Further site-specific analysis here showed that, for the period 2011-2018, an increase in the frequency of extreme anomalies, but not the severity, was observed. This increased frequency of extreme anomalies at Casement Aerodrome is such that the observed probability of being a cluster maximum was 20\% greater here than the upperbound of the credible interval produced by the model using the 1981-2010 data. This increase was also present to a lesser extent at the Phoenix Park and Dun Laoghaire stations (where the observed probability of being a cluster maximum was 7\%  and 2\% respectively above their corresponding upperbounds).


\section*{Acknowledgements}

The authors are grateful to S{\'e}amus Walsh and Sandra Spillane for making the gridded dataset available and for their initial input into this research. This work was supported by the Environmental Protection Agency grant number 2012-CCRP-PhD.3. Andrew Parnell's work was supported by a Science Foundation Ireland Career Development Award grant 17/CDA/4695 and by the Insight Centre for Data Analytics SFI/12/RC/2289.


\bibliographystyle{apalike}
\bibliography{mybib}

\end{document}